\documentclass[twocolumn]{aastex63}

\hypersetup{linkcolor=red,citecolor=blue,filecolor=green,urlcolor=magenta}

\begin{document}

\shorttitle{RR Lyrae $y$-Band Templates}
\shortauthors{Ngeow \& Cheng}

\title{Adding RR Lyrae $y$-Band Template Light Curves to {\tt gatspy}}

\correspondingauthor{C.-C. Ngeow}
\email{cngeow@astro.ncu.edu.tw}

\author[0000-0001-8771-7554]{Chow-Choong Ngeow}
\affil{Graduate Institute of Astronomy, National Central University, 300 Jhongda Road, 320317 Jhongli, Taiwan}

\author{Chia-Yu Cheng}
\affil{Graduate Institute of Astronomy, National Central University, 300 Jhongda Road, 320317 Jhongli, Taiwan}

\begin{abstract}

  We complement the $y$-band template light curves for RR Lyrae to the well-established template light curves in the $ugriz$-band, where the latter have been adopted in the {\tt astroML/gatspy} python package as one of the period-search methods for RR Lyrae. These $y$-band template light curves were constructed based on the $z$-band time series data taken from the Sloan Digital Sky Survey (SDSS), the $y$-band light curves from the Pan-STARRS1 survey, and dedicated $y$-band observations using the Lulin One-meter Telescope for RR Lyrae located in the SDSS Stripe 82 region. These $y$-band template light curves, 9 for the ab-type RR Lyrae and 3 for the c-type RR Lyrae, can be applied in conjunction to the $ugriz$-band template light curves for upcoming sky surveys involving the $y$-band, such as the Vera C. Rubin Observatory Legacy Survey of Space and Time (LSST).
  
\end{abstract}


\section{Introduction}\label{sec1}

It is well-known that the optical light curves for fundamental mode ab-type RR Lyrae (hereafter RRab) exhibit the characteristic ``saw-tooth'' shape with quick rise and slow decline within a pulsation cycle. The optical light curves for first-overtone c-type RR Lyrae (hereafter RRc), on the other hand, tend to be more sinusoidal. Based on repeated observations of 483 RR Lyrae located in the SDSS (Sloan Digital Sky Survey) Stripe 82 region, \citet{sesar2010} derived the RR Lyrae template light curves in the $ugriz$ filters. These template light curves were adopted in the {\tt astroML/gatspy}\footnote{\url{https://github.com/astroML/gatspy}, also see \citet{vdp2016}.} package \citep[hereafter {\tt gatspy};][]{vdp2015} as one of the period-search method.

In coming years, the Vera C. Rubin Observatory Legacy Survey of Space and Time \citep[LSST,][]{lsst2019} will conduct a 10-year time-series synoptic survey in the six $ugrizy$ filters. During early years of LSST operation, the sparsely sampled multi-band light curves would make the period-searching a challenging task \citep[for example, see][]{dic2023}. Since the current available template light curves adopted in {\tt gatspy} only include the five $ugriz$ filters, therefore it is desirable to construct the $y$-band template light curves. In this way, {\tt gatspy} would have a full set of $ugrizy$-band template light curves to be applicable in the era of LSST, or any other similar time-domain surveys involving $y$-band. 

The goal of this work is to construct a set of $y$-band template light curves using the same sample of RR Lyrae located in the SDSS Stripe 82 region. From this sample of RR Lyrae, we first removed those identified as double-mode RR Lyrae \citep[or RRd,][]{varma2024b} or exhibit Blazhko modulations\footnote{A long-term quasi-periodic modulation in amplitude and/or phase in addition to the primary pulsation \citep{blazhko1907}.} in RRab \citep{varma2024a}\footnote{https://etd.lib.nycu.edu.tw/cgi-bin/gs32/ncugsweb.cgi/ccd=\\ DHXaQ5/record?r1=1\&h1=0} and RRc \citep{varma2024}, which left 361 RR Lyrae in our sample. We then retrieved both the SDSS light curves presented in \citet{sesar2010} and the time-series $y$-band data from the Pan-STARRS1 Data Release 2 \citep[PS1DR2;][]{chamber2016,flewelling2020} archive for this sample of 361 RR Lyrae, as detailed in Section \ref{sec2}. We further obtained time-series $y$-band data for a subset of these RR Lyrae using the Lulin One-meter Telescope (LOT, see Section \ref{sec3}). We then combined the available light curves in Section \ref{sec4} to construct the $y$-band template light curves. The conclusion of this work is presented in Section \ref{sec5}.

\begin{figure*}
  \epsscale{1.15}
  \plotone{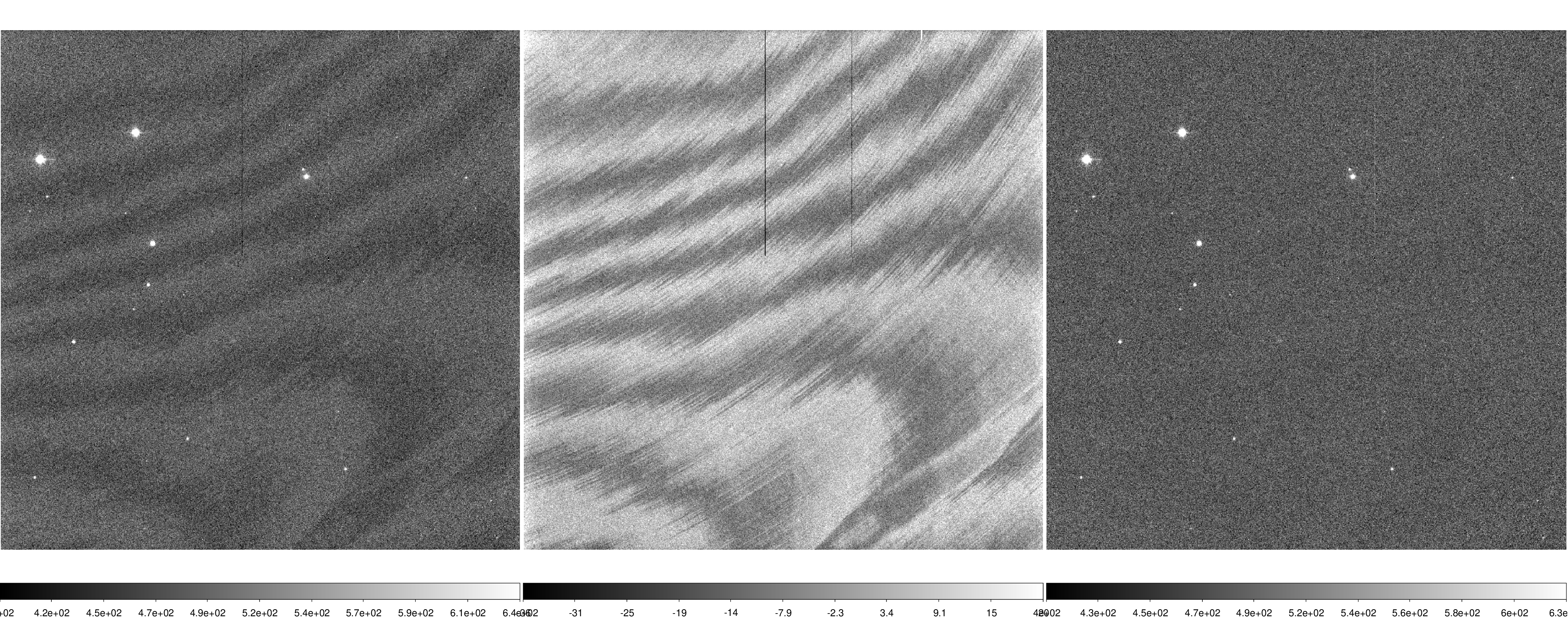}
  \caption{The left panel shows a randomly selected $y$-band reduced image, where fringe patterns can be clearly seen. The middle panel presents the ``fringe-bias'' image produced from the {\tt fringez} package \citep[for more details, see][]{medford2021}, and the right panel demonstrated the fringe patterns have been successfully removed. Specifically, we ran the {\tt fringez-generate} on a subset of images without any bright pixels ($>35000$ ADU; to avoid artifacts from bright stars). The generated fringe models, separately in the $zy$-band, were then applied to all of the reduced images using {\tt fringez-clean}.}
  \label{fig_fringe}
\end{figure*}

\section{Archival Light Curves} \label{sec2}

The near-simultaneous $ugriz$-band observations from SDSS allowed the instantaneous $(g-r)$ colors to be defined. Therefore, the SDSS $z$-band light curves for these RR Lyrae \citep[retrieved from][]{sesar2010} can be transformed to $y$-band using the transformation equation provided in \citet[][their Table 6]{tonry2012}.

To check the reliability of such transformed $y$-band light curves, and to increase the number of data-points per light curves, we have also retrieved the available time-series $y$-band data for this sample of RR Lyrae from the \dataset[PS1DR2]{http://dx.doi.org/10.17909/s0zg-jx37} \citep{p1dr2} archive\footnote{With the API given at \url{https://ps1images.stsci.edu/ps1_dr2_api.html}} available at the Mikulski Archive for Space Telescopes (MAST). For PS1DR2 data, we excluded those measurements with $psf\_Qf\_Perfect<0.95$ \citep{flewelling2020}. The retrieved number of data points per light curves ranged from 14 to 124 and from 1 to 20 for SDSS and PS1DR2, respectively. 

\begin{figure*}
  \epsscale{1.1}
  \plotone{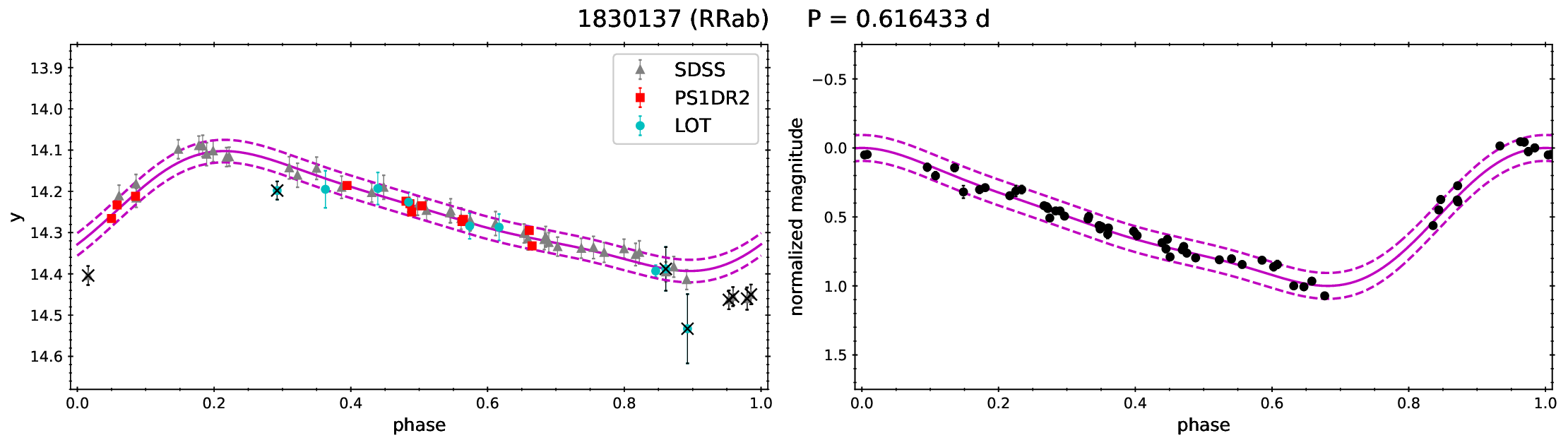}
  \plotone{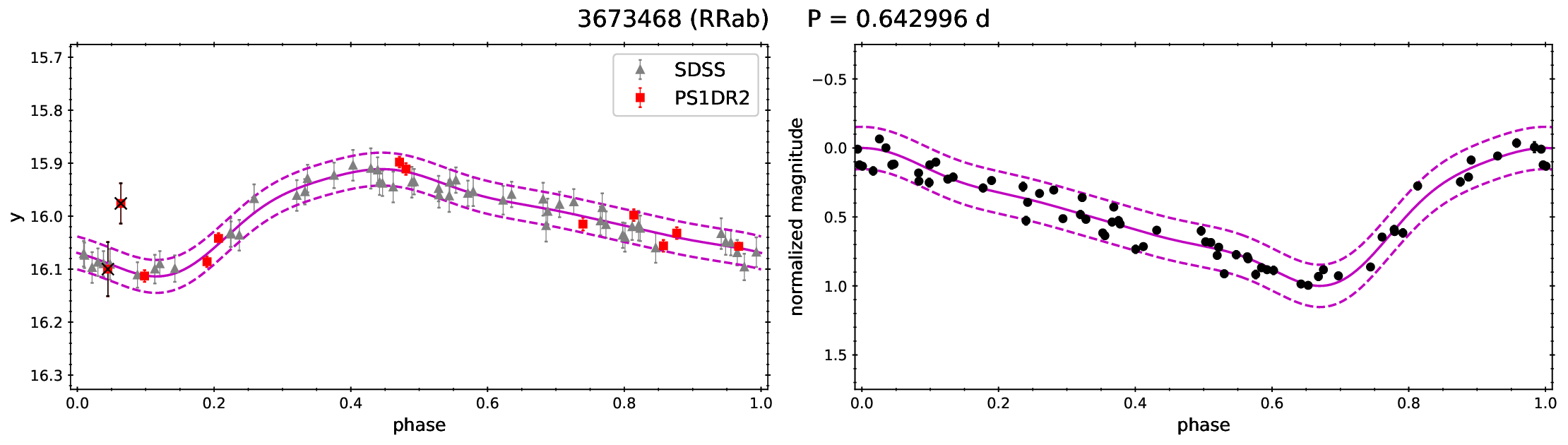}
  \plotone{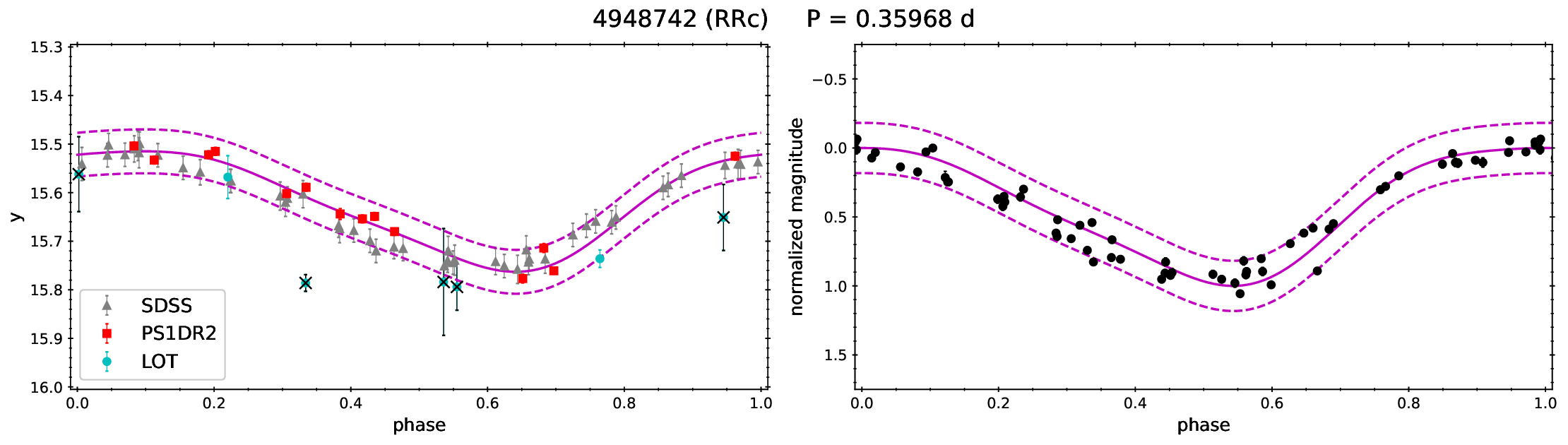}
  \caption{Examples of the $y$-band light curves for two RRab and one RRc stars. The observed light curves are presented in the left panels, where the crosses are rejected data-points (see text for more details). The solid magenta curves are the best-fit Fourier expansion, and the dashed magenta curves represent the $\pm2\sigma$ boundaries. The right panels show the corresponding normalized light-curves, which have been shifted such that the maximum light located at $(0,\ 0)$. }
  \label{fig_nlc}
\end{figure*}

The SDSS light curves turned out to be crucial for the derivation of the $y$-band template light curves. This is because the PS1DR2 $y$-band light curves are too sparse. Even after combined with the LOT data (see next Section), there are only a few of the RR Lyrae with light curve quality sufficient to be used for the template light curve construction. As shown in Figure \ref{fig_nlc}, the transformed SDSS $y$-band light curves are in good agreement with the PS1DR2 light curves.

\section{LOT Time-Series Observations} \label{sec3}

LOT is a $F/8$ Cassegrain telescope located at the Lulin Observatory in central Taiwan. Its main instrument is a Andor iKon-L 936 CCD imager, providing a field-of-view (FOV) of $11.8\arcmin \times 11.8\arcmin$ at a pixel scale of $0.345\arcsec/$pixel. This CCD imager has a low quantum efficiency (QE) at the $y$-band (with QE $\sim 20\%$), implying the exposure time has to be long. Therefore, we only selected RR Lyrae with mean $y$-band magnitudes brighter than $\sim 18$~mag. Depending of the brightness for the targeted RR Lyrae, exposure time varies between 120, 240, 360, and 480 seconds. In total, the LOT queue observations were carried out in 46 full or partial nights from July 2022 to December 2023 (when SDSS Stripe 82 region was visible from the Lulin Observatory). Besides the $y$-band, the selected RR Lyrae were also observed in the $z$-band in $zy$ sequence to properly account for the color-term while calibrating the magnitudes. The commercial $zy$ filters used for our queue observations were purchased from Astrodon. These commercial filters closely resembled the transmission curves of the SDSS and PS1 filters.\footnote{Unfortunately, Astrodon no longer exists and we cannot find its official website. Nevertheless, the transmission curves for the Astrodon filters are available in some vendor's website, e.g.: https://www.firstlightoptics.com/photometry-filters/\\ astrodon-photometrics-sloan-filters.html}

All of the collected CCD images were reduced using the {\tt ccdproc} package \citep{craig2015} written in {\tt python3}. The raw images were bias-subtracted and dark-subtracted, as well as flat-field corrected using either the twilight flats or dome flats. We then discarded a small fraction of ``bad images'', i.e. those images clearly affected by bad weather (e.g. rapid clouds passage), too close to the Moon (and hence exhibit a strong background gradient), or due to tracking problem. For the remaining $zy$-band reduced images, fringe patterns were clearly seen. Hence we employed the {\tt fringez}\footnote{We have to modify the source codes, especially on the filenames, to run the codes on our LOT reduced images.} \citep{medford2021} package to remove the fringe patterns. Figure \ref{fig_fringe} shows an example of reduced image before and after the fringe patterns being removed. After this step, we performed astrometric calibration using the {\tt SCAMP} software \citep[][version 2.10.0]{scamp2006} together with stars in the Two Micron All Sky Survey \citep[2MASS,][]{skrutskie2006} as reference catalog.

For photometric calibration, we followed the approaches described in \citet{ngeow2022}, where the reference stars were queried from the PS1DR2 photometric catalogs with the same selection criteria listed in the Appendix A of \citet{ngeow2022}, except we requested these selection criteria applied to all $grizy$ filters. The photometric calibration were done using the following equations:

\begin{eqnarray}
  z^{PS1} - z^{\mathrm{instr}} & = & ZP_z + C_z (z^{PS1} - y^{PS1}), \\
  y^{PS1} - y^{\mathrm{instr}} & = & ZP_y + C_y (z^{PS1} - y^{PS1}).
  \end{eqnarray}
\noindent In above expressions, $m^{\mathrm{instr}}$ and $m^{PS1}$ represents the instrumental magnitudes and mean magnitudes (in AB system) taken from the PS1DR2 photometric catalogs, respectively. The instrumental magnitudes for the targeted RR Lyrae and the selected PS1DR2 reference stars in the reduced images were measured using the {\tt SExtractor} \citep[][version 2.25.0]{bertin1996} and {\tt PSFEx} \citep[][version 3.17.1]{bertin2011}  software. Specifically, we adopted the point-spread-function (PSF) magnitude, {\tt MAG\_{PSF}}, for measuring the instrumental magnitudes. The zero-point, $ZP_m$, and the color-coefficient, $C_m$, were solved using an iterative $2\sigma$-clipping linear regression. For a pair of $zy$-band images taken in sequence, we have also solved for the color term:

\begin{figure*}
  \epsscale{1.15}
  \plotone{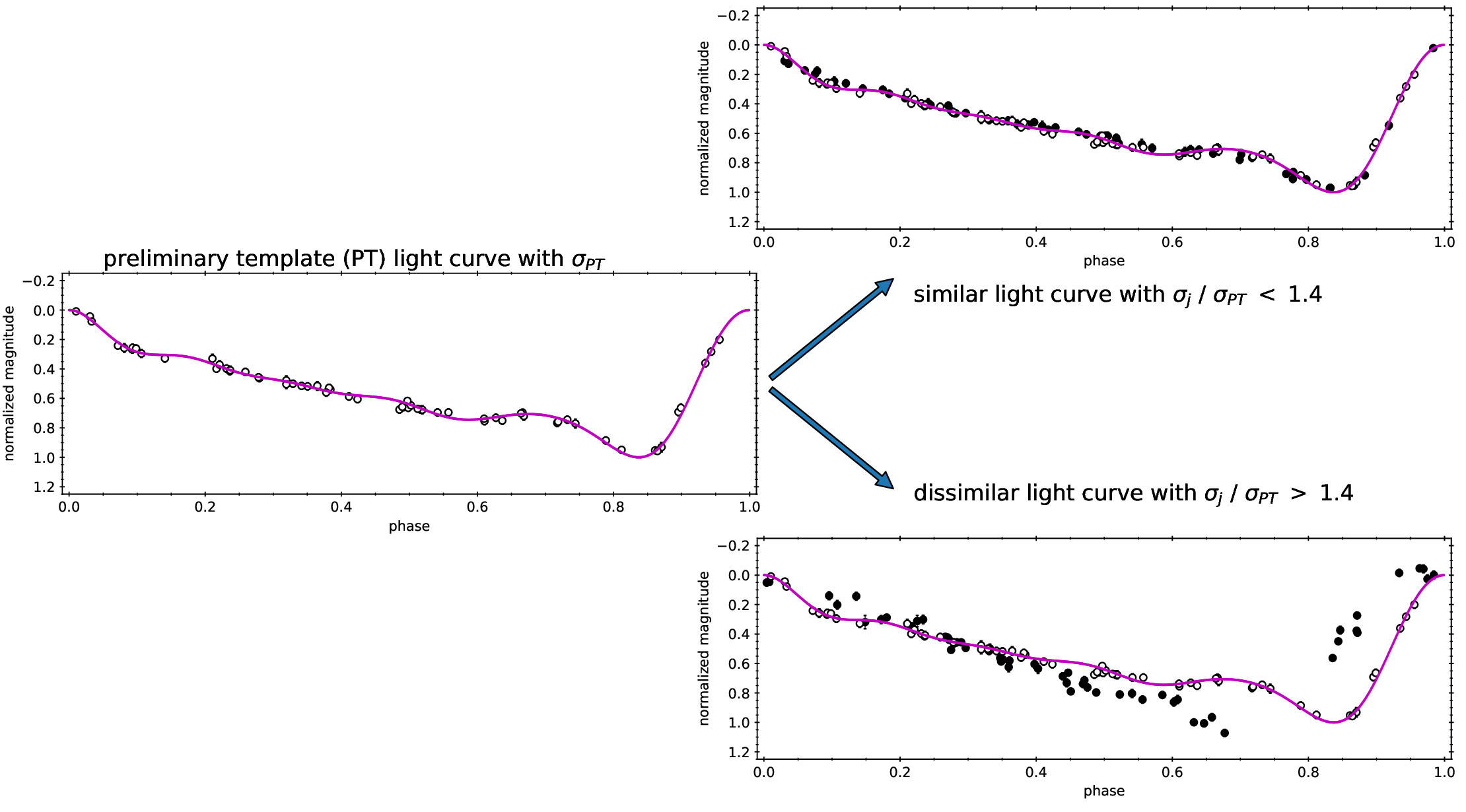}
  \caption{Illustration of selecting a pair of similar light curves. A light curve was selected from a list of RR Lyrae (usually the one with the smallest $\sigma$) as the preliminary template (PT) light curve, which has a light curve dispersion of $\sigma_{PT}$. This PT is shown in the left panel, where the magenta curve is the best-fit light curve to the data-points for this RR Lyrae (open circles). The best-fit light curve (the same magenta curve on the left panel) was then used to fit the $j^{\mathrm{th}}$ observed light curve in the list, represented as filled circles in the right panels, and the light curve dispersion $\sigma_j$ was calculated from the fitting. The top-right panel shows an example of the $j^{\mathrm{th}}$ light curve which is similar to the PT light curve, which satisfied $\sigma_j/\sigma_{PT}<1.4$. In contrast, the bottom-right panel is an example of the $j^{\mathrm{th}}$ light curve that is not similar to the PT light curve. Note that the magenta curves and the open circles in the right panels are same as those in the left panel (i.e. the PT light curve).}
  \label{fig_demo}
\end{figure*}

\begin{eqnarray}
  (z^{PS1} - y^{PS1}) & = & \frac{ZP_z-ZP_y + (z^{\mathrm{instr}} - y^{\mathrm{instr}})}{1-C_z+C_y},
\end{eqnarray}

\begin{deluxetable}{lccc}
  \tabletypesize{\scriptsize}
  \tablecaption{LOT Light Curves for Selected RR Lyrae.\label{tab1}}
  \tablewidth{0pt}
  \tablehead{
    \colhead{Stripe 82 RR Lyrae ID} &
    \colhead{$MJD$} &
    \colhead{$y$} &
    \colhead{$\sigma_y$}
    }
  \startdata
  4099	& 60302.549502 & 16.743 & 0.085  \\
  4099	& 60303.552512 & 17.060 & 0.155  \\
21992	& 59864.699792 & 14.771 & 0.142  \\
21992	& 59892.612176 & 14.729 & 0.127  \\
21992	& 59893.651366 & 14.988 & 0.133  \\
21992	& 59894.608449 & 14.777 & 0.159  \\
$\cdots$ & $\cdots$ & $\cdots$ & $\cdots$ \\
  \enddata
  \tablecomments{This table is available in its entirety in machine-readable form.}
\end{deluxetable}

\noindent and applied the color term back to equation (1) and (2). In total, we have successfully obtained $y$-band data for 134 RR Lyrae. After excluding those measurements with {\tt SExtractor's} $FLAG>0$, there are 1 to 15 data-points per light curves collected from LOT. These light curve data are provided in Table \ref{tab1}.

\section{Constructions of the Template Light Curves} \label{sec4}

\subsection{Preparing the Normalized Light Curves}\label{41}

For each RR Lyrae, we first created a composite $y$-band light-curve by combining their transformed SDSS and the PS1DR2 data, as well as the LOT data if available. During this process, we have also removed data-points with photometric errors larger than 0.05~mag. This step removed 48 faint RR Lyrae, with mean $y$-band magnitudes varying between $\sim19.1$ and $\sim21.3$~mag, in our sample. We further excluded those RR Lyrae with less than 50 data-points on their light curves. The composite light curves for the remaining 215 RRab and 45 RRc were then folded with published pulsation period $P$ \citep[adopted from][]{sesar2010}, and a low-order Fourier expansion \citep[for example, see][]{deb2009} was used to fit the folded light-curve. The best-fit order of the Fourier expansion was determined based on the LOWESS \citep[LOcally WEighted Scatterplot Smoothing,][]{cleveland1979} method \citep[for more details, see][]{ngeow2023}.

The process of fitting the Fourier expansion was iterated twice. In the first iteration, we rejected data-points that are more than $\pm2\sigma$ away from the best-fit light curve, where $\sigma$ is the dispersion of the best-fit light curve. After rejecting the outliers, the remaining data-points were fitted for the second time. The final values of $\sigma$ ranged from 0.02 to 0.30 (mostly for RR Lyrae at the faint end), with a median of 0.07. Once we obtained the best-fitted light curves, we normalized both of the observed and fitted light curves using the amplitudes measured from the best-fitted light curves. At the same time we also determined the phases and magnitudes at the maximum light, and shifted the light curves accordingly such that the phase and magnitude at maximum light for the normalized light curves located at $(0,\ 0)$. Figure \ref{fig_nlc} illustrates this process for two RRab (with and without LOT data) and one RRc stars. During this process, we have also rejected about a dozen of RR Lyrae with $\sigma \gtrsim 0.15$ after visual inspected their normalized light curves.

\begin{figure*}
  \epsscale{1.15}
  \plottwo{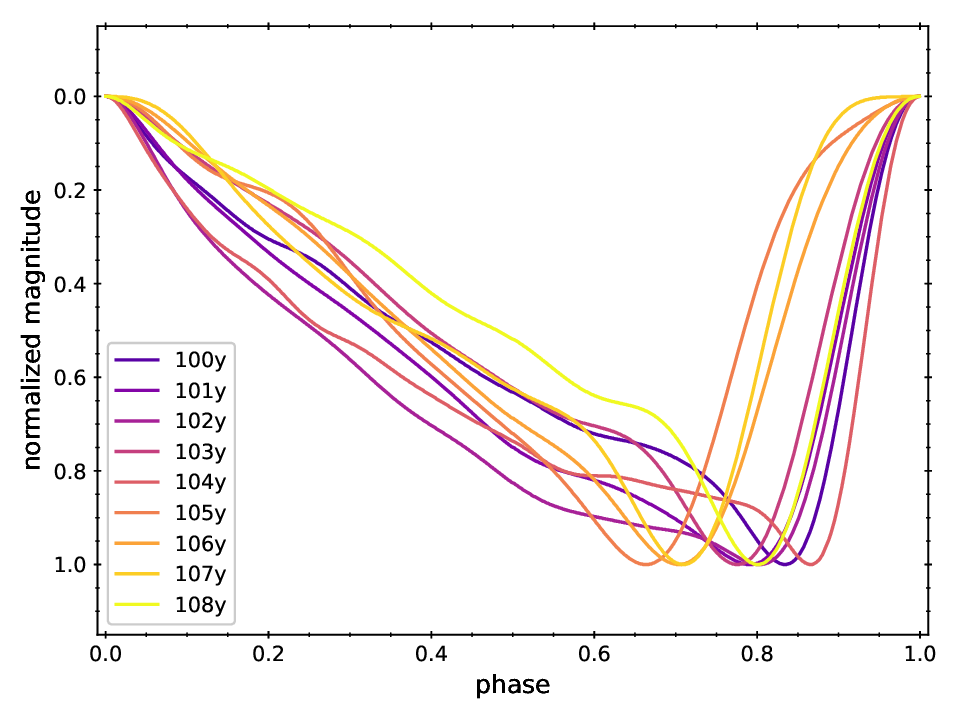}{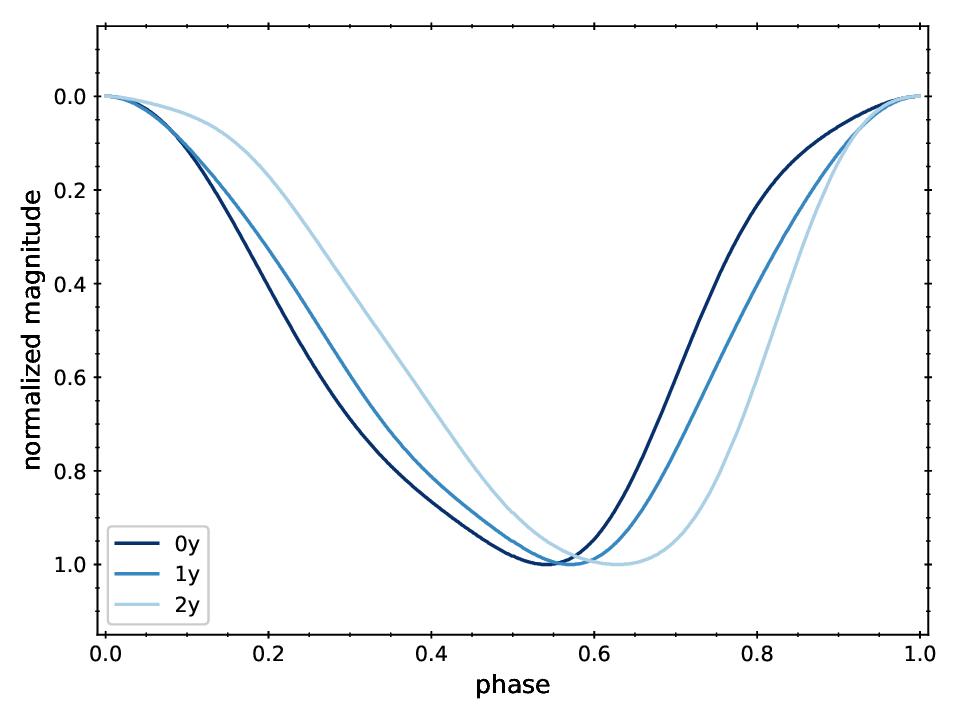}
  \caption{The constructed $y$-band template light curves for RRab (left panel) and RRc (right panel) based on the procedures described in Section \ref{sec4}. These $y$-band template light curves were presented in a number of ASCII files with filenames convention following the $ugriz$-band template light curves given in \citet{sesar2010}.}
  \label{fig_tmpl}
\end{figure*}

\subsection{Constructing the Template Light Curves}

We first assigned a normalized light curve as a preliminary template (PT) light curve, which has a light curve dispersion of $\sigma_{PT}$. We then fit this PT light-curve to the rest of the $j^{\mathrm{th}}$ RR Lyrae in the sample, and calculated the resulted light curve dispersion $\sigma_j$. The normalized light curve for RR Lyrae with the smallest $\sigma_j/\sigma_{PT}$ ratio, as well as $\sigma_j/\sigma_{PT} < 1.4$, would be considered to be similar to the PT (see Figure \ref{fig_demo} for an illustration). Once we identified the $j^{\mathrm{th}}$ normalized light curve was similar to the PT light curve, we combined the data-points from this pair of RR Lyrae. Since the combined light curves have more than 100 data-points, we binned the combined data-points to improve the light curve fittings when using the low-order Fourier expansion (together with the LOESS method). The best-fit Fourier expansion was referred as intermediate template (IT) light curve. These procedures were done separately for the RRab and RRc stars.

It is entirely possible some of the IT light curves were similar and could be combined further. Same as the above procedures, we then assigned an IT light curve as PT light curve, and fit it to the rest of the IT light curves. If the smallest dispersion ratio of the two IT light curves is smaller than 1.2, we considered this pair of RR Lyrae has similar IT light curves and their data-points were combined, binned, and fitted with the low-order Fourier expansion to become a new IT light curve. We iterated these procedures several times until none of the pairs of IT light curves have dispersion ratio smaller than 1.2. Only in a very few cases, we either increased the threshold of the dispersion ratio to 1.3, or not combined the two IT light curves even though they have a dispersion ratio close to unity. All of these decisions were guided via visual inspections. At the end, we constructed 9 and 3 unique $y$-band template light curves for RRab and RRc, respectively. These template light curves are presented in Figure \ref{fig_tmpl}. Note that during the construction of the template light curves, we do not consider the pulsation periods of the RR Lyrae, nor binning the RR Lyrae in period bins.

\begin{figure*}
  \epsscale{1.15}
  \plottwo{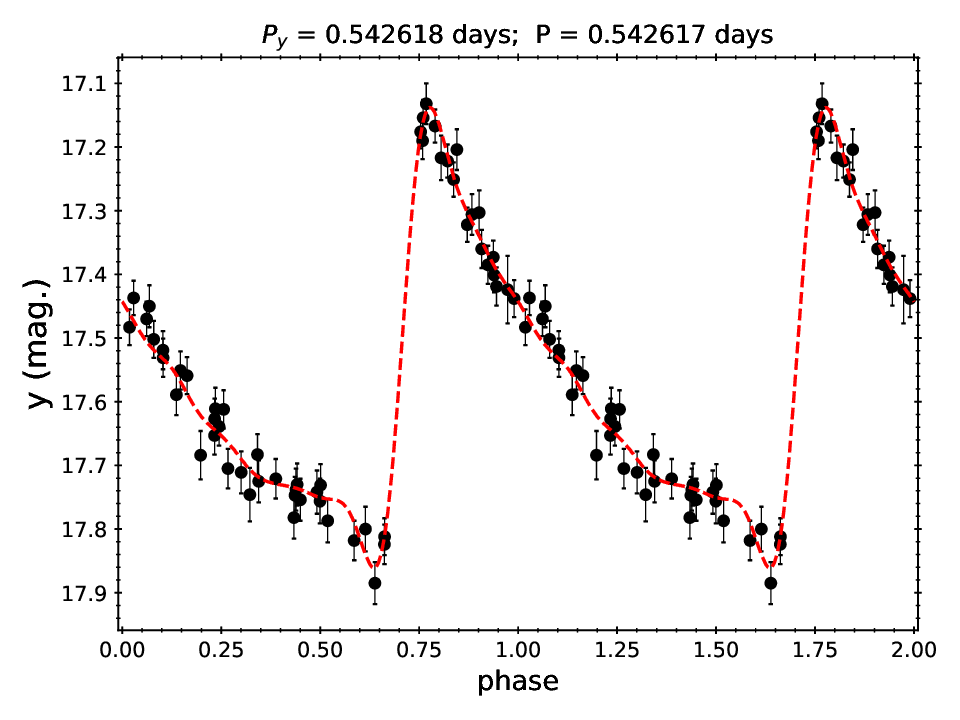}{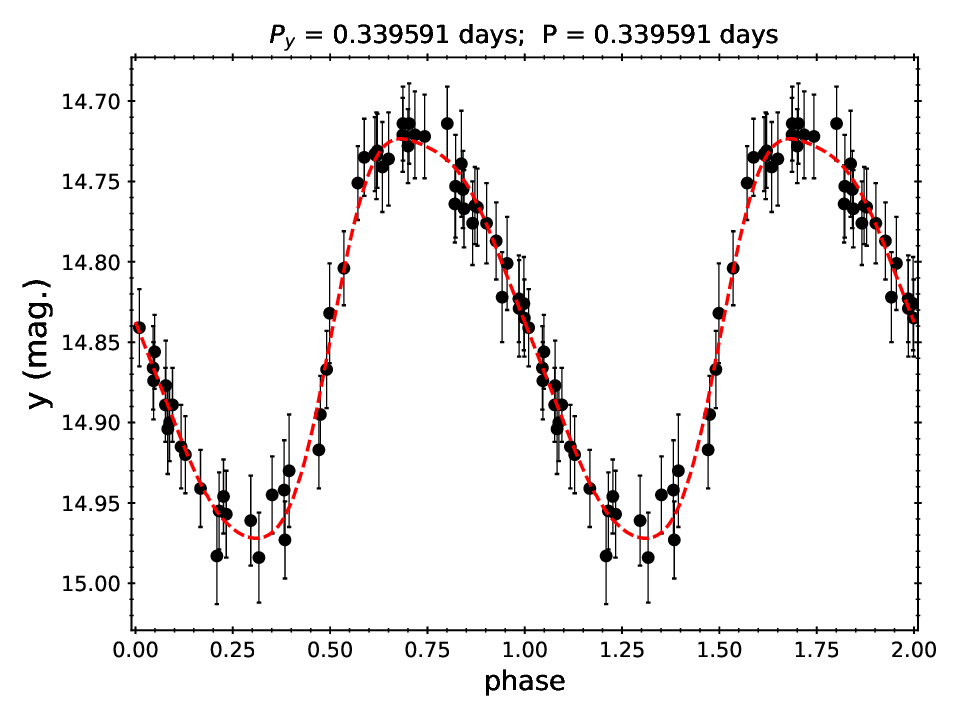}
  \caption{Fitting of the $y$-band template light curves (red dashed curves) to two randomly selected RRab (left panel) and RRc (right panel) stars with {\tt gatspy}. The black points are the transformed SDSS light curve data (see Section \ref{sec2}) for these two RR Lyrae. Periods found by using the $y$-band templates ($P_y$) are in good agreement with the periods given in \citet{sesar2010}.}
  \label{fig_expl}
\end{figure*}

\subsection{Using the Template Light Curves}

The original $ugriz$-band template light curves from \citet{sesar2010} were contained in a gzipped file {\tt RRLyr\_ugriz\_templates.tar.gz},\footnote{This file can be downloaded from \url{https://github.com/astroML/astroML-data/tree/main/datasets}}, which is located in the {\tt astroML\_data/Sesar2010} directory. For convenience, we have included the $y$-band template light curves to the {\tt RRLyr\_ugriz\_templates.tar.gz} file (and not renaming this file), which is available on Zenodo under an open-source Creative Commons Attribution 4.0 license: \dataset[doi:10.5281/ZENODO.14064185]{https://zenodo.org/records/14064185} \citep{ngeow2024}. Hence, users can simply replace this file downloaded from Zenodo to the one in {\tt astroML\_data/Sesar2010} directory, and proceed the usage of {\tt gatspy} as in other $ugriz$ filters. Figure \ref{fig_expl} presents examples of fitting the $y$-band template light curves to two RR Lyrae with {\tt gatspy}.

\section{Conclusions} \label{sec5}

In this work, we constructed the $y$-band template light curves to be included to the {\tt gatspy}. These template light curves were based on the $\sim 250$ RR Lyrae located in the SDSS Stripe 82 region, which have ``good-quality'' light curves (e.g., non Blazhko stars, more than 50 data-points on the light curves with all of them have photometric errors smaller than $0.05$~mag) from the transformed SDSS data, archival PS1DR2 data, and new LOT time-series observations (whenever applicable). These $y$-band template light curve would be useful for the early LSST observations, which is expected to begin very soon, or other sky surveys involving $y$-band on the research using RR Lyrae.

\acknowledgments

We thank the useful discussions and comments from an anonymous referee to improve the manuscript. We are thankful for funding from the National Science and Technology Council (Taiwan) under the contract 109-2112-M-008-014-MY3 and 113-2112-M-008-028. We sincerely thank the observing staff at the Lulin Observatory, C.-S. Lin, H.-Y. Hsiao, and W.-J. Hou, for carrying out the queue observations for this work. This publication has made use of data collected at Lulin Observatory, partly supported by MoST grant 109-2112-M-008-001. This research has made use of the SIMBAD database and the VizieR catalogue access tool, operated at CDS, Strasbourg, France. This research made use of Astropy,\footnote{\url{http://www.astropy.org}} a community-developed core Python package for Astronomy \citep{astropy2013, astropy2018, astropy2022}. This research made use of {\tt ccdproc}, an Astropy package for image reduction \citep{craig2015}.

The Pan-STARRS1 Surveys (PS1) and the PS1 public science archive have been made possible through contributions by the Institute for Astronomy, the University of Hawaii, the Pan-STARRS Project Office, the Max-Planck Society and its participating institutes, the Max Planck Institute for Astronomy, Heidelberg and the Max Planck Institute for Extraterrestrial Physics, Garching, The Johns Hopkins University, Durham University, the University of Edinburgh, the Queen's University Belfast, the Harvard-Smithsonian Center for Astrophysics, the Las Cumbres Observatory Global Telescope Network Incorporated, the National Central University of Taiwan, the Space Telescope Science Institute, the National Aeronautics and Space Administration under Grant No. NNX08AR22G issued through the Planetary Science Division of the NASA Science Mission Directorate, the National Science Foundation Grant No. AST-1238877, the University of Maryland, Eotvos Lorand University (ELTE), the Los Alamos National Laboratory, and the Gordon and Betty Moore Foundation.

\facility{LO:1.0m, PS1, Sloan}

\software{{\tt astropy} \citep{astropy2013,astropy2018,astropy2022}, {\tt ccdproc} \citep{craig2015}, {\tt fringez} \citep{medford2021}, {\tt gatspy} \citep{vdp2015}, {\tt PSFEx} \citep{bertin2011}, {\tt SCAMP} \citep{scamp2006}, {\tt SExtractor} \citep{bertin1996}}



\end{document}